\documentclass[aps,prl,preprint,showpacs,groupedaddress]{revtex4}
\begin{document}
\title{EFFECT OF STIMULATED RADIATION PROCESSES ON FORMATION OF TRACKS AND NANOCAVITIES IN SOLIDS}
\author{Fedor V.Prigara}
\affiliation{Institute of Microelectronics and Informatics,
Russian Academy of Sciences, \\21 Universitetskaya, Yaroslavl
150007, Russia} \email{fprigara@imras.yar.ru}
\date{\today}

\begin{abstract}
It is shown that the average diameter of tracks produced by swift
heavy ions in metals is determined by the size of elementary
resonator. The last is introduced in the theory of thermal
radiation accounting for the stimulated radiation processes. Due
to the inverse density gradient, radiation produced by electronic
excitations is locked in the region of track which is formed.
Similar effects are applied to the process of nanocavity formation
in semiconductors during ion irradiation.
\end{abstract}
\maketitle

Up to now, no convincing mechanism for the formation of damage
tracks produced in metals by super-high energy ions is proposed.
The mechanism of Coulomb burst is not relevant in the case of
metals because of high rate of neutralization \cite{1}. The
mechanisms based on the interaction of electrons with phonons and
other collective modes which can exist in the periodic crystals
(e.g., plasmons) seem to be non-adequate due to the violation of
space periodicity in the region of track formation.

Super-high energy ions lose their energy in solids by producing
the dense electronic excitations \cite{1,2}. Here we show that
thermal radiation produced by electrons plays a very important
role in the process of track formation in solids and determines
the average diameter of tracks produced by super-high energy ions
in metals.

The local melting of crystal lattice in the region of track
formation makes it possible to apply the theory of thermal
radiation proposed for gaseous media \cite{3} to the radiation
produced by electronic excitations.

It was shown recently \cite{3} that thermal emission has a
stimulated character. According to this conception thermal
emission from non-uniform gas is produced by an ensemble of
individual emitters. Each of these emitters is an elementary
resonator the size of which has an order of magnitude of mean free
path \textit{l} of photons

\begin{equation}
\label{eq1}
l = \frac{{1}}{{n\sigma} }
\end{equation}

\noindent
where \textit{n} is the number density of particles and $\sigma $ is the
absorption cross-section.

The emission of each elementary resonator is coherent, with the wavelength

\begin{equation}
\label{eq2}
\lambda = l,
\end{equation}

\noindent
and thermal emission of gaseous layer is incoherent sum of radiation
produced by individual emitters.

An elementary resonator emits in the direction opposite to the direction of
the density gradient. The wall of the resonator corresponding to the lower
density is half-transparent due to the decrease of absorption with the
decreasing gas density.

The condition (\ref{eq2}) implies that the radiation with the wavelength $\lambda $
is produced by the gaseous layer with the definite number density of
particles \textit{n} .

The condition (\ref{eq2}) is consistent with the experimental
results by Looney and Brown on the excitation of plasma waves by
electron beam (see \cite{4,5}). The wavelength of standing wave
with the Langmuir frequency of oscillations depends on the density
as predicted by equation (\ref{eq1}). The discrete spectrum of
oscillations is produced by the non-uniformity of plasma and the
readjustment of the wavelength to the length of resonator. From
the results of experiment by Looney and Brown the absorption
cross-section for plasma can be evaluated.

The product of the wavelength by density is weakly increasing with the
increase of density. This may imply the weak dependence of the size of
elementary resonator in terms of the wavelength upon the density or,
equivalently, wavelength.

Because of the high opacity of metals and the inverse radial density
gradient, thermal radiation produced by electron excitations is locked in
the region of track formation. The characteristic diameter of channel, in
which the radiation is locked, is the size of an elementary resonator. The
results of experiment by Looney and Brown suggest that the size of an
elementary resonator, \textit{d}, differs by the numerical factor of 4 from
the mean free path of photons:

\begin{equation}
\label{eq3}
d = 4l = 4/n\sigma .
\end{equation}

Substituting in the last equation $n \cong 10^{22}cm^{ - 3}$ and
$\sigma \cong 10^{ - 15}cm^{2}$, we find that $d \cong 4nm$. Such
is the average diameter of tracks measured in alloys irradiated by
super-high energy ions \cite{1}.

Thermal radiation locked in the region of track formation is
heating electrons supporting the dense electronic excitations. Due
to the violation of space periodicity in the track region, it is
more relevant to suggest that the energy and momentum of electrons
transfer to the lattice atoms in the process of individual
electron-ion or electron-atom collisions, rather than by the
interaction of electrons with collective modes existing in a
crystal lattice. The region of track formation has a fluid-like
structure and radial distribution function \cite{1}.

The proposed mechanism can also contribute to the process of
nanocavity formation in semiconductors during keV ion irradiation
\cite{6}, since the inverse radial density gradient locking the
thermal radiation here is also present. The average diameter of
nanocavity has an order of magnitude of the size of an elementary
resonator. The results of Ref.6 suggest that the radiation
pressure play an important role in the nanocavity formation
process. In accordance with equation (\ref{eq2}), the
characteristic wavelength, $\lambda $, of radiation has an order
of magnitude of the size of an elementary resonator. In the case
of nanocavities created in Ge by 5 keV Xe ion irradiation,
$\lambda \cong 10nm$\cite{6}, hence $\hbar \omega \cong 0.1keV$,
where $\hbar $ is the Planck constant, and $\omega $ is the
frequency of radiation. If the radiation temperature is $T_{r}
\cong \hbar \omega \cong 0.1keV$, then the radiation pressure is
$P_{r} = \alpha T_{r}^{4} \cong 0.2GPa$. Thus the radiation
pressure in nanocavity is comparable with the estimated
equilibrium pressure \cite{6}.

To summerize, the role of thermal radiation in the processes of track and
nanocavity formation in solids irradiated by high energy ions cannot be
ignored.

\end{document}